\documentstyle[epsf,rotate]{article}
\begin{document}

\begin{center}

{
\Large
    {\bf The Influence of the Initial Mass Function on Populations of X-ray
Binaries After a Burst of Star Formation}

\vskip 0.3cm

}

Popov S.B., Lipunov V.M., Prokhorov M.E., Postnov K.A.\\

Sternberg Astronomical Institute\\
Moscow State University\\
119899, Moscow, Universitetskii prospekt, 13\\
e-mail: polar@xray.sai.msu.su; lipunov@sai.msu.su;
  mike@sai.msu.su;  moulin@sai.msu.su\\

Astronomy Reports V. 42, No. 1 pp. 29-32 (1998).\\
Astronomicheskii Zhurnal V.75. N1. P. 35-39\\

Recieved: April 20, 1997

\end{center}

    In this article we use "Scenario Machine" - the population synthesis
simulator- to calculate the evolution of the populations of the selected
types of X-ray sources after the starformation burst with the total mass
in binaries $1.5\cdot10^6 M_{\odot}$ during the first 20 Myr after the burst.
Sources of the four types were calculated:
transient sources- accreting  neutron stars with  Be- stars;
accreting neutron stars in pair with  supergiants; Cyg X-1-like sources-
 black holes with  supergiants; superaccreting
black holes. We used two values of the $\alpha - coefficient$ in the
mass-function: 2.35 (Salpeter's function) and 1.01 ("flat spectrum").
The calculations were made for two values of the upper limit of the
mass-function: 120 and 30 $M_{\odot}$.
For the flat spectrum the number of sources of all types significantly
increased.
Decreasing of the upper mass limit below the critical mass
of a black hole formation increase (for the "flat spectrum")
the number of transient sources with neutron stars up to $\approx 300$.
We give approximating formulae for the time dependence of source numbers.

\pagebreak

\section{Introduction}

    From the point of view of stellar evolution, objects undergoing bursts
of star formation are of special interest. One example of such objects are
Wolf-Rayet galaxies [1]. Considering the only evolution of single stars,
Contini et al. [2] suggested that some observational propertiesof the galaxy
Mrk 712 could be understood if the initial mass function were substantially
different from the standart value. Specifically, they proposed a ``flat''
mass spectrum with index $\alpha$=1 in place of a Salpeter mass spectrum
with index $\alpha$=2.35, with an upper mass limit of $120 M_{\odot}$.
Later, Schaerer [3], also studying the evolution only of single stars,
showed that the observations can also be explained using a Salpeter initial 
mass function.

In [4], we investigated a burst of star formation under the conditions at
the center of the Galaxy, and showed that massive X-ray binary systems are
sometimes better indicators of the presence and age of starbursts than the
optical properties of populations of single stars (color indices, etc.). One
of our goals here is to draw attentionof population-synthesis specialists to
the necessity of taking into account the possible effects of synthesis
parameters on the evolution of populations of binary stars. We consider the
evolution of binary systems after burst of star formation with two initial
mass functions and two upper mass limits.

\section{The Model}

 Our `Scenario Machine'' population synthesis simulator was first described
in [5], and recently, a detailed description was given in the review [6]
(see also [4]). Therefore, here, we will only make note of the parameters
used in the calculations.

We considered two power-law initial mass functions, with indices
$\alpha$ 2.35 É 1.01.

\begin{equation}
   \frac{dN}{dM} \propto M^{-\alpha}
\end{equation}

 We calculated $10^7$ binary systems in each run of the program. The lower
mass limit for the more massive component was taken to be $0.1 M_{\odot}$
in the normalization and $10 M_{\odot}$ in the computations. Less massive
stars do not evolve sufficiently over a time $<2\cdot10^7$ years to produce
neutron stars and black holes (unless their mass is appreciably increased by
accretion). The results were normalized to a total mass of binary stars in
the starburst of $1.5\cdot10^6 M_{\odot}$. We present here calculations for
upper mass limits of $120  M_{\odot}$ and  $30  M_{\odot}$. We chose the
rather exotic upper mass limit of $30  M_{\odot}$ for the following reasons.
In population synthesis modeling, usually two parameters are varied -- the
slope and upper mass limit of the initial mass function (see [11], where, in
fact, an upper limit of $30  M_{\odot}$ is used). We chose oue smaller value
for the upper mass limit to be less than the critical mass for the formation
of a black hole so that it would be possible to study this situation. We
took the relative mass distribution for the binary components to be flat,
i.e., systems with components with similar and with very different masses
were equally probable.

 We took the major axes of the systems to be in the interval from 10 to
$10^7 R_{\odot}$, with a logarithmically uniform distribution. The magnetic
fields of neutron stars were assumed not to decay; this is natural, given
the relatively short evolution period investigated (see [10] for a
discussion of the field decay). A black hole was formed when the mass of the
pre-supernova star was more than 35 $M_{\odot}$, 70\% of the initial mass
became part of the black hole. The Oppenheimer- Volkov lomit was chosen to
be  2.5 $M_{\odot}$. We did not use the ``enhanced'' mass loss rate proposed
by the Geneva group (see, for example, [3,9]). We will take this possibility
into account in future papers.

 In anisotropic supernovae explosions, the compact object (neutron star or
black hole) acquired an additional velocity. These velocities had random
directions and a distribution of amplitudes corresponding to that observed
for radiopulsars [7], with a characteristic value of 200 km/s (see [8] for a
discussion of our choice of a lower characteristic velocity than proposed in
[7]).

\section{Results}

  We present our results for four types of sources: (1) X-ray transients
associated with neutron star -- Be star systems; (2) accreting neutron star
-- supergiant systems (one of the types of binary systems that give birth to
X-ray pulsars); (3) systems with a black hole accreting at a rate above the
Eddington limit (SS 433 may be such an object); and (4) accreting black hole
-- supergiant systems (one example of this type of objects is Cyg X-1).

%

 The calculated evolution of the number of sources for the two values of the
initial mass function index are presented in Figs.1 (upper mass limit 30
$M_{\odot}$) and 2 ((upper mass limit 120 $M_{\odot}$). We have not smoothed
the curves, and their ``noisiness'' is due to statistical fluctuations (we
calculated $10^7$ binary systems for each mass interval and then normalized
the results). X-ray transients were the most numerous type of sorce because
of the long lifetimes of these systems at this stage (the figures show the
{\it observable} number of systems at each time). It stands to reason that
we have no information about whether a transient is in an active stage at
any given moment in time.

 When the upper mass limit is lowered, the number of sources with black
holes decreases and the number with neutron stars increases. As we decrease
the upper mass limit below the critical mass for the formation of black
holes, sources with black holes may not disappear entirely, since they can
form if the masses of neutron stars are increased to the Oppenheimer -
Volkov limit by accretion. This is most likely when the accretion is
supercritical (and depends on the assumptions about the supercritical
accretion stage), as was clear from our calculations. For this reason, the
number of sources such as SS 433 rose with time when the upper mass limit
was  $30 M_{\odot}$, rather than undergoing the sharp decrease observed for
the upper mass limit of  $120 M_{\odot}$, when the vast majority of black
holes are formed from massive stars in the first 3-4 million years after the
starburst. Systems such as Cyg X-1 are not shown for the upper mass limit of
$30 M_{\odot}$, since not a single system of this type formed over the
calculation time.

 We present approximation formulas for the convenience in determining the
number of various types of sources formed in starbursts with arbitrary
masses.
In these expressions, the time $t$ is in million of years. In the case of a
Salpeter ($\alpha = 2.35$) initial mass function with an upper limit
$M_{up}=120 M{\odot}$, we find for X-ray transients at times from 5 to 20
million years after the starburst

\begin{equation}
    N(t)=-0.14\cdot t^2+5.47\cdot t -14.64   .
\end{equation}

 For superaccreting black holes at times from 4 to 20 million years

\begin{equation}
    N(t)=\frac{2.2}{t-3.05}.
\end{equation}

 For sources such as Cyg X-1, we have at times from 4 to 20 million years

\begin{equation}
    N(t)=\frac{4.63}{t-2.9}  .
\end{equation}

 We find for binary systems consisting of an accreting neutron star and a
supergiant at times from 5 to 20 million years

\begin{equation}
    N(t)=2.12\cdot 10^{-4} \cdot t^3 -9.6\cdot 10^{-3}
    \cdot t^2 +0.13\cdot t -0.47.
\end{equation}

 For a flat initial mass function with an upper mass limit $M_{up}=120
M{\odot}$, we obtain for X-ray transients at times from 3 to 7 million years 

\begin{equation}
    N(t)=-8.9\cdot t^2 +1.2\cdot 10^2 \cdot t -3 \cdot 10^2,
\end{equation}

\noindent
and at times from 7 to 20 million years

\begin{equation}
    N(t)=-2.8\cdot t +1.2\cdot 10^2.
\end{equation}

 For superaccreting black holes at times from 4 to 20 millino years, we have

\begin{equation}
    N(t)=\frac{39.97}{t-3.17}       .
\end{equation}

 For sources such as Cyg X-1 at times from 4 to 20 million years,

\begin{equation}
    N(t)=\frac{58.44}{t-3.08}         .
\end{equation}

 Finally, for accreting neutron star - supergiant binaries, we find
for times from 5 to 20 million years

\begin{equation}
    N(t)=1.45\cdot 10^{-3}\cdot t^3 -5.96\cdot 10^{-2}\cdot t^2+
     0.74\cdot t -2.41.
\end{equation}

\section{Conclusion}

In [4], we showed that the absolute and relative number of massive baniry
systems with black holes and neutron stars can serve as a good indicator of
the age of a burst of star formation. Here, we considered the evolution of
four populations of close binary systems with neutron stars or black holes
after a starburst with a total mass of $1.5\cdot10^6 M_{\odot}$ in binary
systems for the cases of Salpeter ($\alpha$=2.35) and flat ($\alpha$=1.01)
initial mass functions, following the evolution for $2\cdot 10^7 $ years
after the starburst. We included the exotic upper mass limit of  $30
M_{\odot}$ in our study in order to model populations with a strongly
reduced number of black holes (note that this situation could also come
about in the presence of a strong stellar wind associated with a high mass
loss rate).

 The transition to the initial mass function index $\alpha=1.01$ proposed in
[2] leads to a significant increase in the number of X-ray sources. We
expect the number of single accreting neutron stars and black holes, and
also the number of other binary systems, to increase as well, but we have
not considered these objects here. 

 Lowering the upper mass limit below the critical mass for the formation of
black holes leads to the virtually complete disappearance of sources with
black holes, but increases the number of transient sources with neutron
stars (this increase is especially strong for a flat initial mass function,
to  $\approx 300$ neutron star systems). Thus, this change of the initial
mass spectrum brings about a substantial change in the luminosity of the
galaxy at standart and hard X-ray energies. In other words, a burst of star
formation with a flat initial mass function should be accopmanied by a large
X-ray flux. The absence of a large X-ray luminosity suggests that the
initial mass spectrum for the given starburst was not flat.

\vskip 0.3cm
ACKNOWLEDGMENTS

We thank the referee, whose comments significantly improved the paper. This
work was supported by INTAS 93-3364 and the Russian Foundation for Basic
Research  95-02-06053a. S.B.P. thanks the Soros Educational Program for
1995-1997.

\pagebreak

{
\Large{References}
}

\noindent
1. Conti P.// Astrophys. J. 1991.  V. 377. P. 115.

\noindent
2. Contini T., Davoust E., Considere S.//Astron. and Astrophys. 1995.
V. 303. P. 440.

\noindent
3. Schaerer D.// Astrophys. J. 1996. V. 467. L17.

\noindent
4. Lipunov V.M., Ozernoy L.M., Popov S.B., Postnov K.A., Prokhorov M.E.//
Astrophys. J. 1996. V. 466. P. 234.

\noindent
5. Kornilov V.G., Lipunov V.M. // AZh  1983. V. 60. p. 284.

\noindent
6. Lipunov, V.M., Postnov, K.A., Prokhorov, M.E.//
Astroph. and Space Phys. Rev. 1996. V.9. part 4.

\noindent
7. Lyne A.G., Lorimer D.R.// Nature. 1994. V. 369. P.127.

\noindent
8. Lipunov V.M., Postnov K.A., Prokhorov M.E.// Astron. and Astrophys. 1996.
V. 310. P. 489.

\noindent
9.  Schaller G.,  Schaerer D.,  Meynet G.,  Maeder A.//
 Astron. and Astrophys. Supp. 1992. V.96. P.269

\noindent
10. Konenkov D. Yu. , Popov S.B.// Astronomy Letters 1997. V.23. p.569.

\noindent
11. Perez-Olia D.E., Colina L.// Mon. Not. Royal  Astron. Soc. 1995. V. 277.
P. 857.

The text was translated by D. Gabuzda

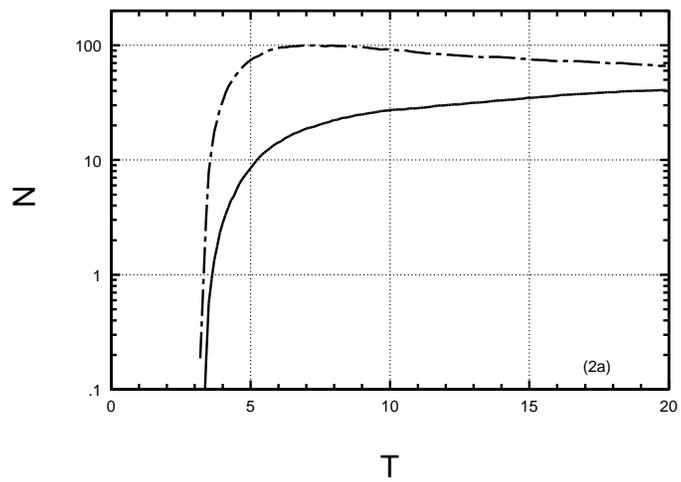
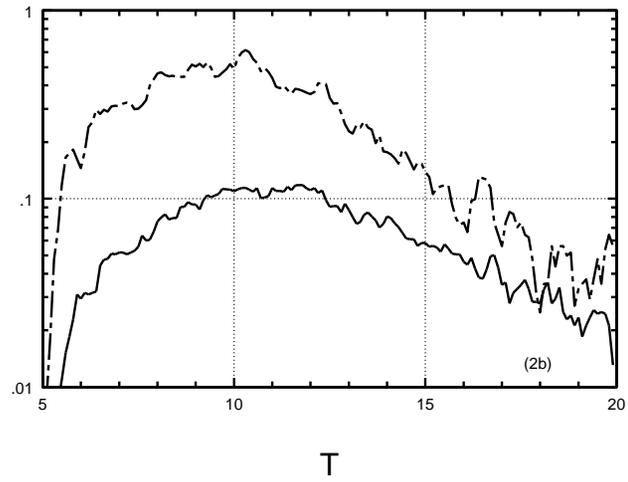
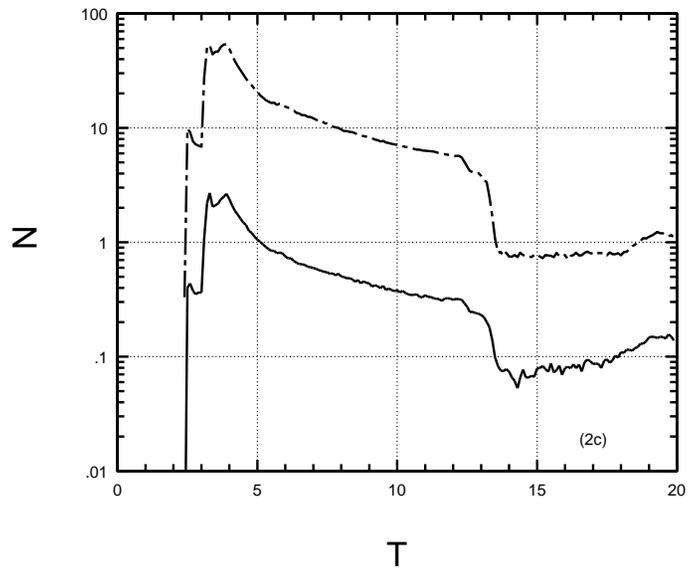
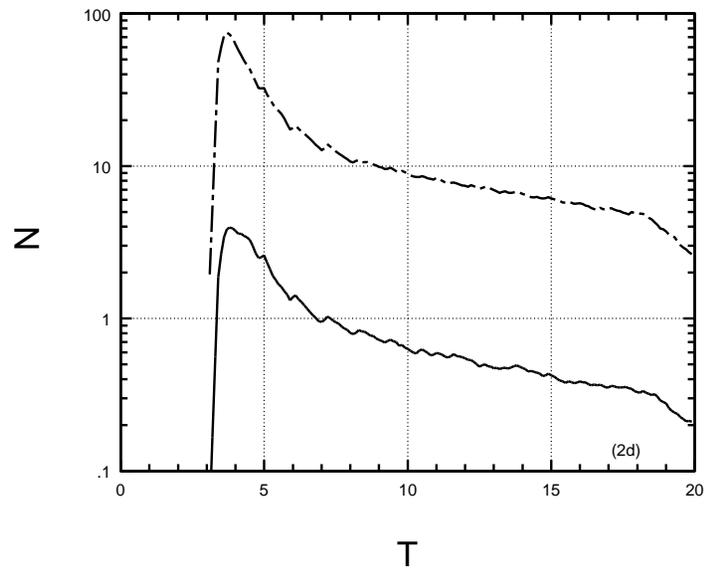
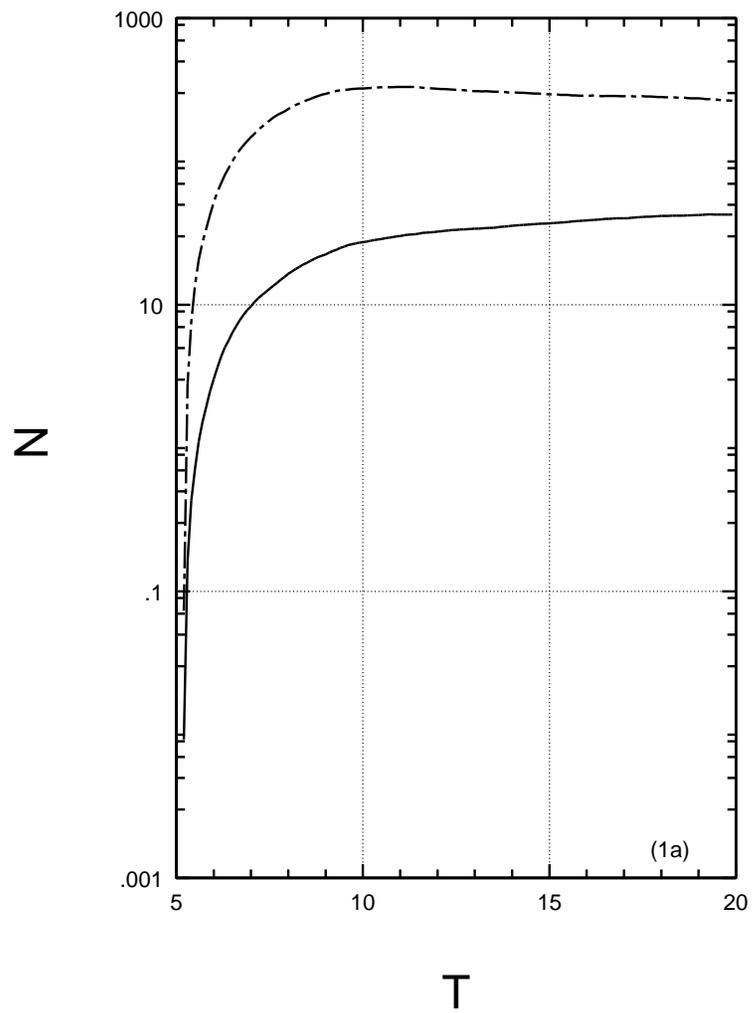
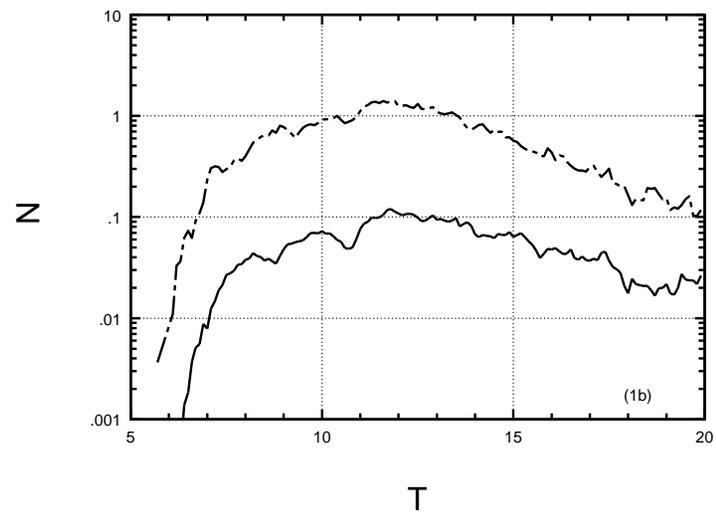
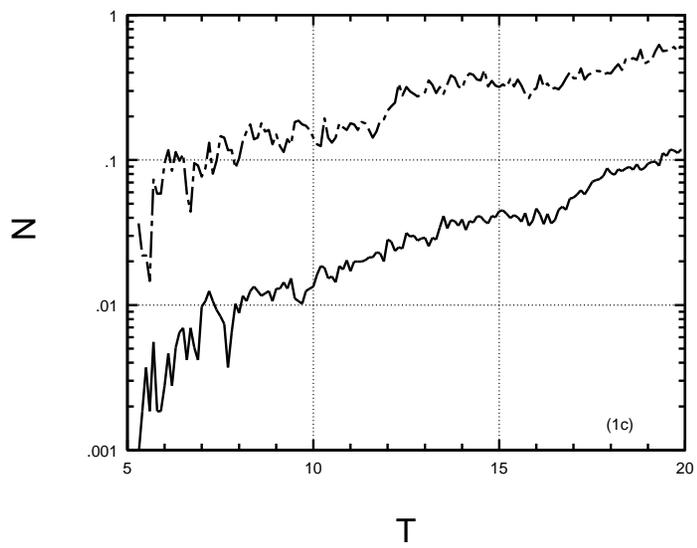
\end{document}